\documentclass[fleqn,usenatbib]{mnras}

\usepackage{newtxtext,newtxmath}
\usepackage[T1]{fontenc}

\DeclareRobustCommand{\VAN}[3]{#2}
\let\VANthebibliography\thebibliography
\def\thebibliography{\DeclareRobustCommand{\VAN}[3]{##3}\VANthebibliography}

\usepackage{graphicx}	% Including figure files
\usepackage{amsmath}	% Advanced maths commands

\title[Synthetic cycle for anti-Hale regions]{Synthetic solar cycle for active regions violating the Hale's polarity law}

\author[A. Zhukova et al.]{
A. Zhukova,$^{1}$\thanks{E-mail: anastasiya.v.zhukova@gmail.com}
A. Khlystova,$^{2}$\thanks{E-mail: hlystova@iszf.irk.ru}
V. Abramenko$^{1}$\thanks{E-mail: vabramenko@gmail.com}
and D. Sokoloff$^{3,4,5}$\thanks{E-mail: sokoloff.dd@gmail.com}
\\
$^{1}$Crimean Astrophysical Observatory, Nauchny, 298409, Crimea\\
$^{2}$Institute of Solar Terrestrial Physics, Siberian Branch, Russian Academy of Science, Irkutsk 664033, Russia\\
$^{3}$Department of Physics, Lomonosov Moscow State University, Moscow 119992, Russia\\
$^{4}$Moscow Center of Fundamental and Applied Mathematics, Moscow 119992, Russia\\
$^{5}$Pushkov Institute of Terrestrial Magnetism, Ionosphere and Radio Wave Propagation, Russian Academy of Science, Troitsk 108840, Moscow, Russia
}

\date{Accepted XXX. Received YYY; in original form ZZZ}

\pubyear{2022}

\begin{document}
\label{firstpage}
\pagerange{\pageref{firstpage}--\pageref{lastpage}}
\maketitle

\begin{abstract}
%not more than 250 words no references.

Long observational series for bipolar active regions (ARs) provide significant information
about the mutual transformation of the poloidal and toroidal components of the global solar magnetic field.
The direction of the toroidal field determines the polarity of leading sunspots in ARs in accordance
with the Hale's polarity law.
The vast majority of bipolar ARs obey this regularity, whereas a few percent of ARs have the opposite sense of polarity (anti-Hale ARs).
However, the study of these ARs is hampered by their poor statistics.
The data for five 11-year cycles (16--18 and 23,24) were combined here to compile a synthetic cycle of unique time length and latitudinal width.
The synthetic cycle comprises data for 14838 ARs and 367 of them are the anti-Hale ARs.
A specific routine to compile the synthetic cycle was demonstrated.
We found that, in general,  anti-Hale ARs follow the solar cycle and are spread throughout the time-latitude diagram evenly,
which implies their fundamental connection with the global dynamo mechanism and the toroidal flux system.
The increase in their number and percentage occurs in the second part of the cycle,
which is in favour of their contribution to the polar field reversal.
The excess in the anti-Hale ARs percentage at the edges of the butterfly diagram and near
an oncoming solar minimum (where the toroidal field weakens) might be associated with strengthening of the influence of turbulent convection and magnetic field fluctuations on the arising flux tubes.
The evidence of the misalignment between the magnetic and heliographic equators is also found.

\end{abstract}

\begin{keywords}
dynamo -- Sun: activity -- Sun: magnetic fields
\end{keywords}

%%%%%%%%%%%%%%%%%%%%%%%%%%%%%%%%%%%%%%%%%%%%%%%%%%

%%%%%%%%%%%%%%%%% BODY OF PAPER %%%%%%%%%%%%%%%%%%

\section{Introduction}

Solar activity is widely associated with the mutual transformation of the poloidal and toroidal
components of the global magnetic field during each solar cycle (SC).
However, many essential details of this mechanism are far from being understood quantitatively and qualitatively. 
The transformation is hidden in the depths
of the convective zone near the tachocline and is inaccessible for direct observations.
Despite the success in helioseismology, photospheric observations remain
the dominant source of information about the processes taking place in the solar interior.
In particular, long observational series spanning for more than a century exist for sunspot groups.

In every SC, it is the direction of the toroidal field that determines the features of active regions (ARs) in the N- 
and S-hemispheres of the photosphere on the Sun.
The most obvious characteristic of ARs is the leading sunspot polarity.
The Hale polarity law states that the leading sunspots (leaders) in bipolar sunspot groups in the Northern hemisphere
during even/odd cycles have the same positive/negative polarity, whereas the polarity of the leaders
in the Southern hemisphere show the opposite tendency \citep{Hale19}.
This key empirical fact had been the motivation and underlies the magnetic cycle models, starting from the pioneering ones
by \citet{Parker55, Babcock61, Leighton64}.
The vast majority of bipolar ARs satisfy the established regularity, although a small part of them happen to show the converse 
behavior which can be termed as the Anti-Hale polarity law. 

This was first emphasized by \citet{Hale25} and repeatedly confirmed in further studies
\citep[see e.g.][]{Richardson48, Smith68, Vitinsky86, Wang89, Khlystova09, McClintock14,
Abramenko18, Li18, Munoz-Jaramillo21}.
So far, occurrence of reverse-polarity ARs has no generally accepted interpretation.
Note that not all ARs on the Sun display the bipolar structure.
Among others, there are many unipolar and multipolar ARs where the Hale polarity law is not
always applicable \citep[see e.g.][]{Abramenko18, Zhukova20GA, Abramenko21}.
Anti-Hale ARs constitute only about $3 \%$ of total number of ARs.

Studies of ARs violating the Hale polarity law (anti-Hale ARs) play a significant role
in deepening our understanding about some essential aspects of solar activity in such as the role of the turbulent component 
of dynamo on different scales \citep{Stenflo12, Sokoloff15}, formation of magnetic flux tubes \citep{LopezFuentes00, 
LopezFuentes03, Nelson13} and especially about the intricasies involved with the polar field reversals \citep{Hazra17, 
Karak18, Mordvinov19, Mordvinov22}. Also the knowledge gained in such studies help us to 
predict the characteristics of upcoming solar cycle which has a very important practical implication for the rapidly 
growing space and satellite industry \citep{Mouradian93, Nagy17, Nagy19}.

A reverse-polarity orientation is often found in ARs with dominating  $\delta$-configuration
of the magnetic field.
Such ARs are known for their high flaring productivity and initiation of strong geomagnetic events and therefore emphasize 
their importance in the influence of space weather
\citep{Smith68, Knizhnik18, Abramenko21, Kashapova21}.

%Observation on the reverse-polarity ARs and dynamo modeling are widely used

Unfortunately, the study of such groups is hampered by their poor statistics
\citep{Zhukova20Sola}, and this is an obstacle for conducting some essential tests (for instance, to analyse
the time-latitude distribution of anti-Hale groups in detail).
Meanwhile, the famous Maunder butterfly diagram is one of the main observational patterns
of solar activity \citep[e.g.][]{Hathaway15, Usoskin17}.

The aim of this work is to clarify the features of the temporal and time-latitude distribution
of anti-Hale ARs and to find possible relation to the mechanisms of solar magnetic field evolution.
%and discuss a possible relationship between the areas of ARs-``violators'' concentration and physical mechanisms.}
For this purpose we offer a new observational test based on the combination of data of several solar cycles and
compilation of a synthetic cycle. This will allow us to improve the statistics of anti-Hale groups.
The synthetic cycle that consists of 14838 ARs of five SCs (16--18 and 23,24), including 367 anti-Hale groups,
is also used to clarify the issue of mutual position of the magnetic and heliographic equators.

\section{Data}

As a source of reverse-polarity bipolar groups(anti-Hale ARs) of SCs 23 and 24, we used
the Catalog of bipolar active regions violating the Hale polarity law from 1989-2018 (BARVHL),
and its prolongation (2019-2020), which can be accessed at\, \url{http://sun.crao.ru/databases/catalog-anti-hale}.
The verification of ARs in this catalog was carried out in accordance with the criteria
for identifying the anti-Hale groups based on a thorough study of complex ambiguous cases
\citep{Zhukova20Sola}.
The main criteria assume that an anti-Hale AR is a group with sunspots (at least pores) of both
polarities, which are linked with pronounced magnetic connection verified by a presence of UV-loops.
These criteria were also applied in the present study when dealing with ARs of the period from 1925 to 1958.
The Summary of Mount Wilson magnetic observations of sun-spots (SMWMO) is the source of the information
about reverse-polarity ARs for this time interval.
The Summary was published in the Publications of the Astronomical Society of the Pacific.
The SMWMO data are available in digital form at \url{http://iopscience.iop.org/journal/1538-3873}.

We also used the data from modern solar instruments,
namely magnetic field data from SOHO/MDI \citep{Scherrer95}, SDO/HMI \citep{Scherrer95} and EUV data
from SOHO/EIT \citep{Moses97}, SDO/AIA \citep{Lemen12}.
We relied on the Debrecen Photoheliographic Data (DPD)
at \url{http://fenyi.solarobs.csfk.mta.hu/DPD} \citep{Baranyi16}
and the Mount Wilson observatiory (MWO) drawings at \url{ftp://howard.astro.ucla.edu/pub/obs}.

We also used the Royal Greenwich Observatory and United States Air Force/National Oceanic and Atmospheric Administration 
Solar Region
Summary (USAF/NOAA SRS) at \url{http://solarcyclescience.com/activeregions.html}
to get the data for all ARs that appeared on the solar disk from 1919 to 1959 and from 1989 to 2021.

\section{Synthetic cycle compilation}

To overcome statistical limitations discussed in Introduction and to explore the time-latitude distribution of anti-Hale groups, we used the idea of combining data from several cycles.

The averaged period of the Schwabe cycle is known to be close to 11 years.
Although the cycles differ slightly in their duration and amplitude, they have approximately the same shape.
As it was mentioned by \citet{Hathaway15}, ``an average cycle can be constructed by stretching and contracting
each cycle to the average length, normalizing each to the average amplitude''.
Implementation of this idea for cycles 1-23 is shown in Fig. 26 in \citet{Hathaway15}.

Using their approach and taking into account the latitudinal positions of ARs, we developed the method of creating a synthetic cycle.
For this purpose,  for several full cycles we needed information on all ARs including anti-Hale groups, that emerged on the disk from the beginning to the end of each cycle.
Therefore, we used data for those cycles where the information on anti-Hale ARs is available.
The elongated catalog BARVHL (1989-2020) provided us with data on the cycles 23 and 24.
The SMWMO data from 1925 to 1958 allowed us to add the information on cycles 16, 17, and 18.
During these cycles in the SMWMO bipolar ARs with reverse polarity are marked, which allowed us to compile a special database of verified anti-Hale ARs for this time interval.

All this allowed us to operate with the data for five solar cycles.

\subsection{Determination of a hosting cycle for each AR}
\label{sec:det}

Compiling a correct database for each individual cycle is a challenge.
A simple segregation of ARs by time is not suitable
due to the overlapping periods during each solar minimum when both low-latitude ARs of the old cycle
and high-latitude precursors of the new cycle appear on the disk simultaneously.
First of all, we have to determine which cycle a given AR belongs to. This should be performed
for each AR.

\begin{figure}

\includegraphics[width=\columnwidth]{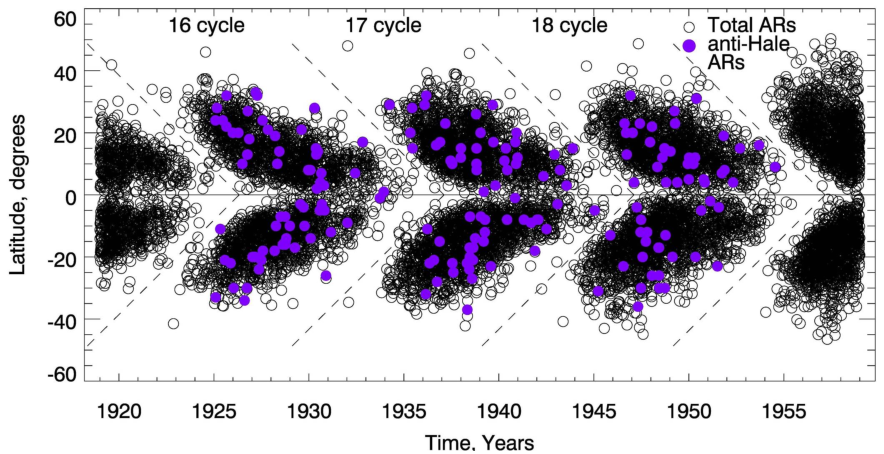}
\includegraphics[width=\columnwidth]{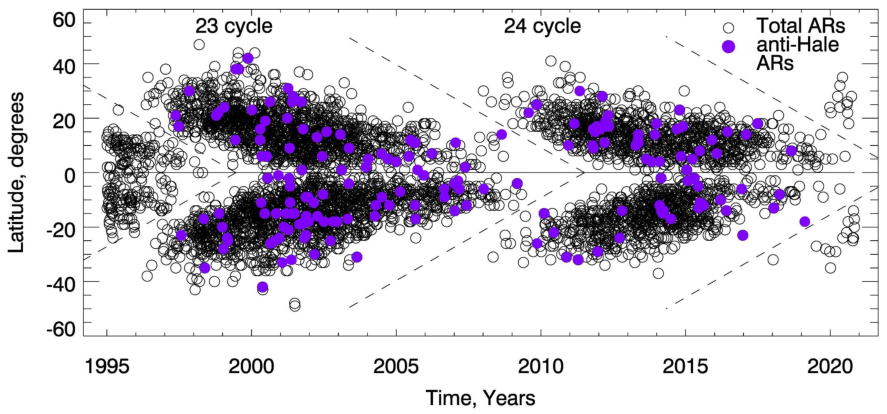}

\caption{Time-latitude distribution of sunspot groups: cycles 16, 17, 18 (top panel); cycles 23, 24 (bottom panel).
Anti-Hale ARs are marked with filled violet circles, overlapping all ARs
that are represented as unfilled black circles. Border lines between the cycles are shown as dotted lines.}
\label{fig:bound}
\end{figure}

To define the solar cycles' boundaries as accurately as possible, we used the technique suggested by \citet{McClintock14}.
According to their approach, a boundary between adjacent cycles is a line with a slope of 1/60
(during sixty days the latitude changes by one degree).
The point of intersection of this line with the equator was adopted so that the high-latitude ARs of a new polarity must be located to the right from this border line, as it is shown in Fig.\,\ref{fig:bound}.

\subsection{Mutual adjustment of the time-latitude diagrams of the five cycles}
\label{sec:fitt}

\begin{figure}

\includegraphics[width=\columnwidth]{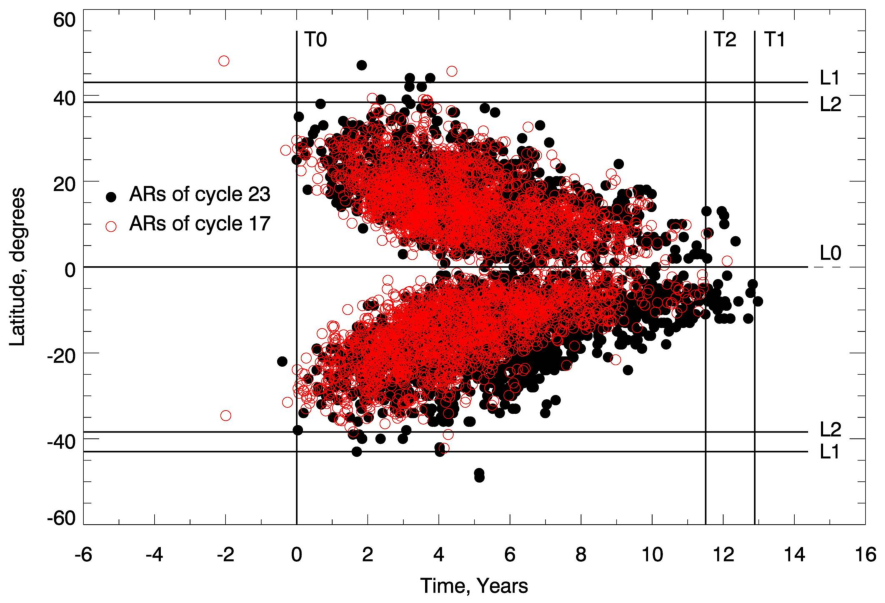}
\includegraphics[width=\columnwidth]{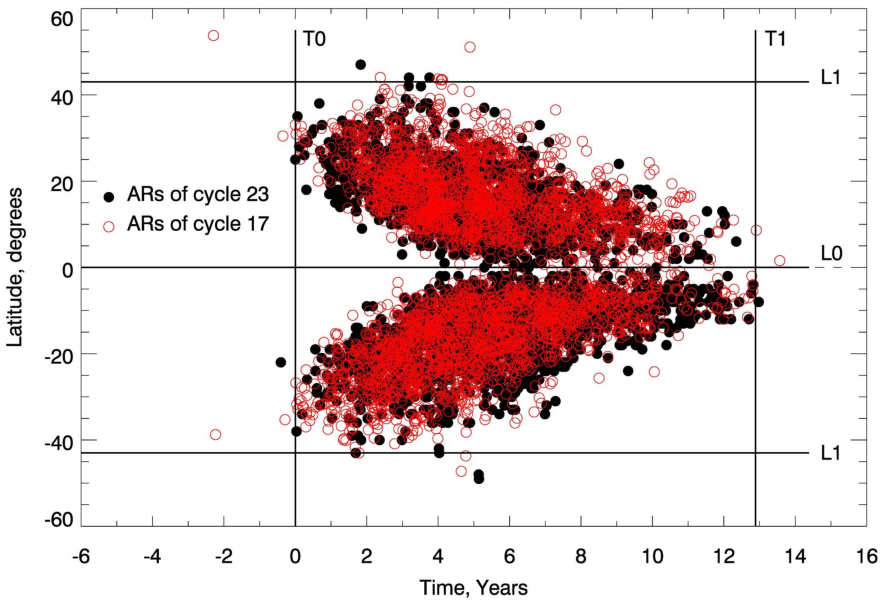}

\caption{Overlapping of two SCs before (top panel) and after (bottom panel) the linear transformation.
ARs of the basic SC 23 are plotted with filled black circles, ARs of the SC 17 are overplotted with unfilled red circles.
The lines T1 and L1 show the most reasonable fit for the SC 23, whereas the  narrower cycle 17 was bounded with lines
T2 and L2. The ratios L1/L2 and T1/T2 were used as coefficients for linear stretching of the date from SC 17  to match the two cycles
in latitude and time. Bottom - the overlap between cycle 23 and stretched cycle 17.}
\label{fig:trans}
\end{figure}

Time-latitude diagrams of the cycles that are overplotted do not coincide.
The length and width of the butterfly wings are different for different cycles,
as it is shown in the top panel in Fig.\,\ref{fig:trans} for SCs 23 and 17.

\begin{figure}

\includegraphics[width=0.95\columnwidth]{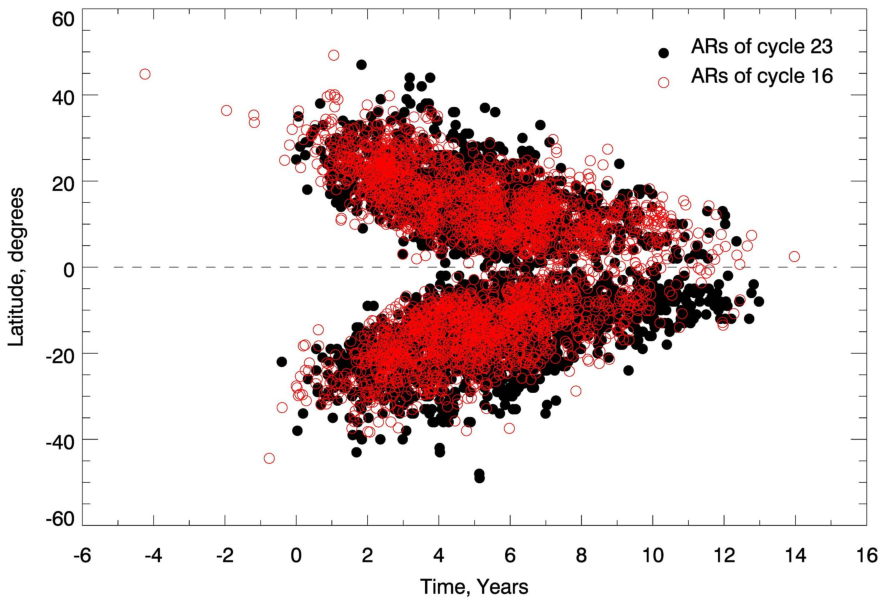}
\includegraphics[width=0.95\columnwidth]{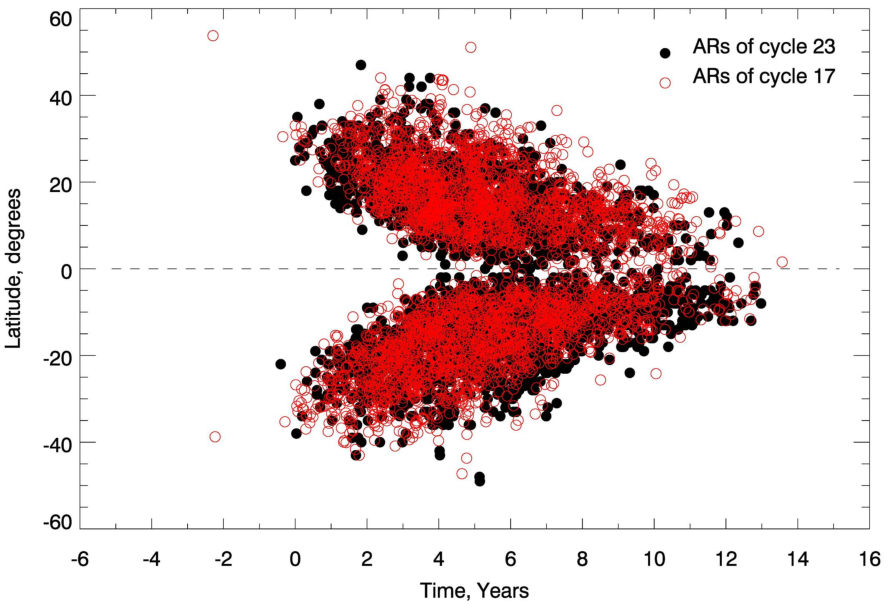}
\includegraphics[width=0.95\columnwidth]{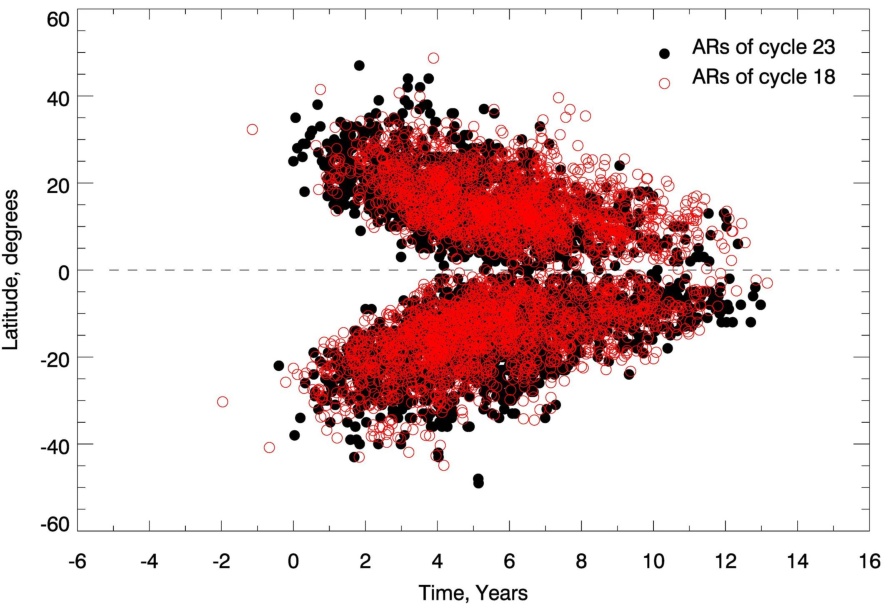}
\includegraphics[width=0.95\columnwidth]{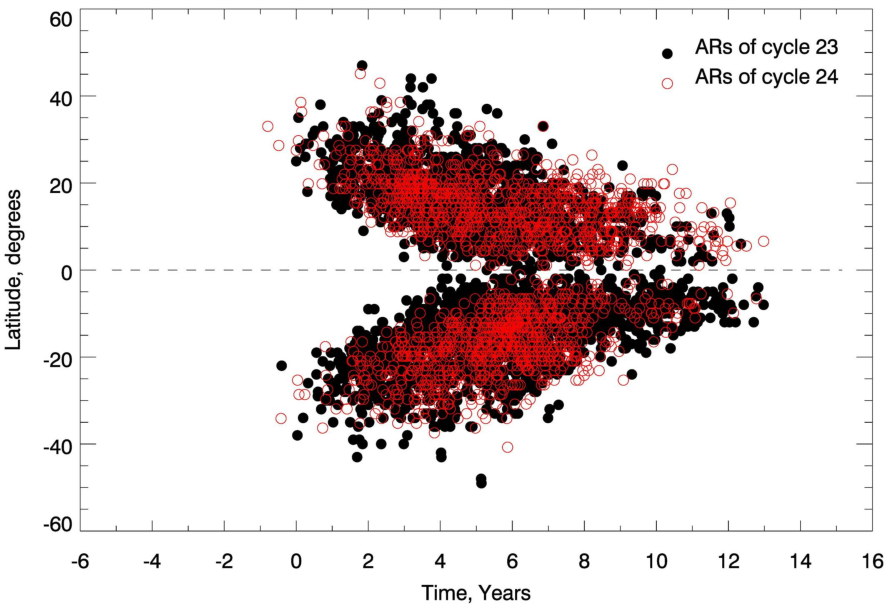}

\caption{Fitting of SCs 16, 17, 18, 24 data to SC 23.
Notations are the same as in Fig.\,\ref{fig:trans}.
}
\label{fig:fitt}
\end{figure}

To achieve the matching of butterfly wings, we performed a linear transformation in both time and latitude coordinates.
The SC 23 was taken as the basic one because the ARs in this cycle are distributed over the largest area on the time-latitude
diagram.
The start point for each cycle, T0 in Fig.\,\ref{fig:trans}, was adopted as the moment when their ARs begin to emerge regularly
(at least two or three ARs per month).
The equatorial line was accepted as the natural latitude divider and marked as L0.
The lines T1 and L1 bound the SC 23. While selecting their location, we neglected single distant data points.
In the similar way, the lines T2 and L2 outline the SC 17.
The ratios T1/T2 and L1/L2 were used as the extension coefficients for the coordinates of the ARs of SC 17.
As a result of this procedure the butterfly wings for the SC 23 and the SC 17 occupy approximately the same area
on the time-latitude diagram, see the bottom panel in Fig.\,\ref{fig:trans}.
The same routine was performed for the rest three cycles.
The graphic results are summed up in in Fig.\,\ref{fig:fitt}, whereas the transformation coefficients are shown
in Table\,\ref{tab:coeff}.

\begin{table}
	\centering
	\caption{Transformation coeffitients.}
	\label{tab:coeff}
	\begin{tabular}{lccccr}
		\hline
		Solar cycle & 16 & 17 & 18 & 23 & 24\\
		\hline
		Time coeffitient & 1.07 & 1.12 & 1.15 & 1.0 & 1.1\\
		Latitude coeffitient & 1.07 & 1.12 & 1.0 & 1.0 & 1.1\\
		\hline
	\end{tabular}
\end{table}

\section{Variations of anti-Hale ARs along the synthetic cycle}

\begin{figure}

\includegraphics[width=\columnwidth]{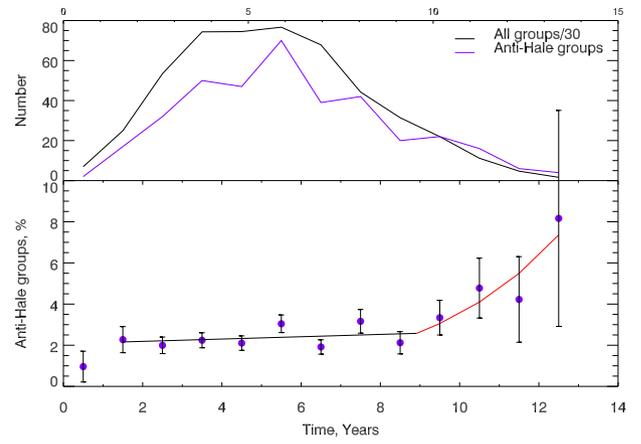}

\caption{Top panel: temporal variations of all (thick black line) and anti-Hale
(thin violet line) ARs in the synthetic cycle.
Bottom panel: yearly values of the percentage of anti-Hale ARs (violet circles).
Black line and red curve represent the fitting with linear and exponential functions respectively.}
\label{fig:synthperc}
\end{figure}

Advantages of using the synthetic cycle is the enhanced statistics of rare anti-Hale ARs and a possibility to study
the population of ARs during a solar minimum without contamination caused by the high-latitude ARs belonging to the next cycle.
A distribution of all 14838 ARs along the years of the synthetic cycle is shown in Fig.\,\ref{fig:synthperc} (top panel)
with a black curve inorder to give a better visual presentation and the data were normalized by 30.
In the same figure, the distribution of 367 anti-Hale ARs along the systhetic cycle is shown (the violet line).
The distribution of all ARs is rather smooth, although the presence of two maxima \citep[first
emphasized by][and observed in the most SCs]{Gnevyshev63} is quite distinguishable.
A total duration of the synthetic cycle is about thirteen years.
Note that the period of the sporadic singular occurence of the earliest new-cycle precusors is not displayed here.

Anti-Hale ARs (thin violet line) show overall correspondence with the cycle and demonstrate
a pronounced peak in the second maximum of the synthetic cycle.
It is consistant with our previous results \citep[an increase of irregular ARs, including anti-Hale ones,
was found in the second maxima of SCs 23, 24 by][]{Abramenko18, Zhukova20GA}.

The anti-Hale ARs relative number shown in the bottom panel in Fig.\,\ref{fig:synthperc}), on the contrary, is cycle-independent
(with the exception of the late declining phase of the cycle and solar minimum).
The percentage is about $2.5 \%$ during the cycle, and we approximated it by the linear function
with the coefficient $0.06$.
During the first year of the cycle, the fraction of anti-Hale ARs is negligible (about $1 \%$).
During the late declining phase (years $11-13$), the fraction of anti-Hale groups seems to be growing.
Unfortunately, the interval of this growth is very short and error bars are large, and it is not possible
to make a reliable fit.
The red curve in the bottom panel in Fig.\,\ref{fig:synthperc} only indicates the trend of an increase.

\begin{figure}

\includegraphics[width=\columnwidth]{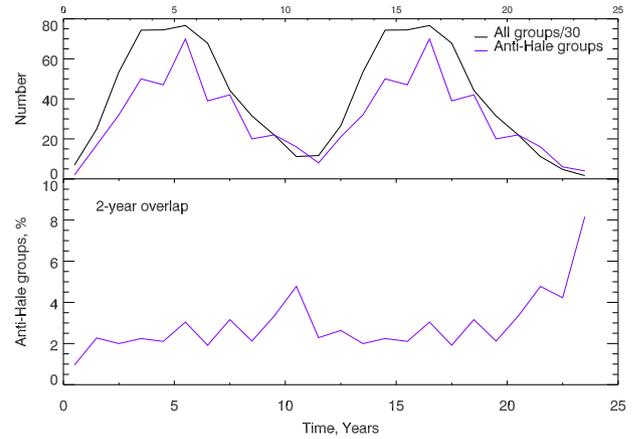}

\caption{Temporal variations of ARs in the doubled synthetic cycle with overlapping intervals
one (top paired panels) and two (bottom paired panels) years.
 Notations are the same as in Fig. 1. for each pair of panels}
\label{fig:double}
\end{figure}

To simulate the adjunction of cycles, we combined two identical synthetic cycles with different
overlap intervals.
The doubled synthetic cycle with an overlapping interval of two years is shown in Fig.\,\ref{fig:double}.
We found a pronounced peak (about $5 \%$) in the solar minimum.
The result for an overlapping interval of one year is similar.
Note that such overlapping intervals are consistent with an observed period of simultaneous occurrence
of ARs of the adjacent cycles.
However, with the three-year overlapping interval the peaks were not observed.

This finding is in accordance with previous observations for SCs 20-23 by \citet{McClintock14} and
SCs 15-17 by \citet{Sokoloff15}, who reported an enhancement in the anti-Hale ARs percentage during the solar minima.
Our thorough analysis of the original data on each anti-Hale AR
from the BARVHL catalog and SMWMO data showed that the peaks in anti-Hale percentage
occur immediately before or at the moment when new-cycle groups start to appear.
It is interesting that the vast majority of anti-Hale groups
belongs to old cycles during every minimum, and the precursors of
each new cycle are regular ARs following the Hale polarity law.

\section{Time-latitude distribution of ARs in the synthetic cycle}

\begin{figure}

\includegraphics[width=\columnwidth]{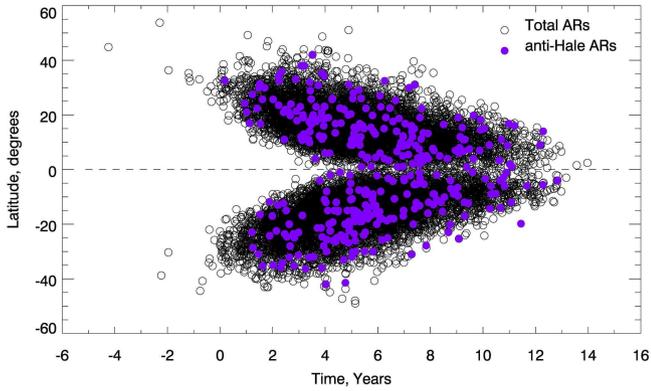}

\caption{Time-latitude distribution of 14838 ARs (open black circles) along the synthetic cycle.
367 anti-Hale ARs (violet circles) are plotted over all observed ARs.}
\label{fig:butt}
\end{figure}

All ARs of the synthetic cycle are presented in Fig.\,\ref{fig:butt} as a time-latitude (butterfly) diagram as open black circles.
The anti-Hale ARs are overplotted with filled violet circles.
One can see that the anti-Hale ARs are spread nearly uniformly over the diagram.
%, which implies the ubiquitous presence of magnetic field fluctuations in their formation.

\begin{figure}

\includegraphics[width=\columnwidth]{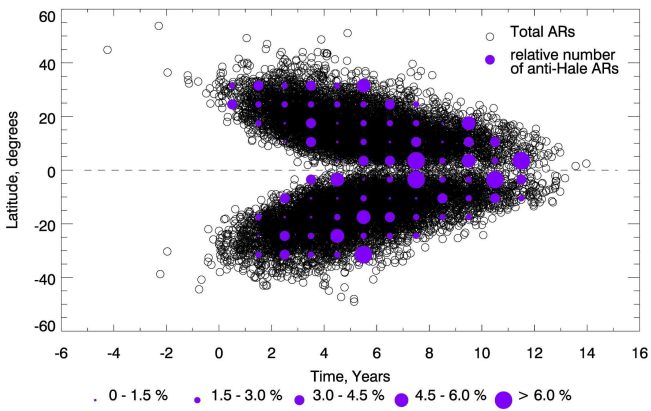}

\caption{Time-latitude distribution of the relative number of anti-Hale ARs (filled violet circles) of the synthetic cycle.
The size of each circle is proportional to the relative number (the scale is shown
under of the panel).
The background diagram is the same as in Fig.\,\ref{fig:butt}. }
\label{fig:relbutt}
\end{figure}

The time-latitude distribution of the relative number of anti-Hale ARs is presented in Fig.\,\ref{fig:relbutt}.
The entire space of the butterfly diagram was binned into equal intervals in time (one year) and latitude
(seven degrees).
The relative number of anti-Hale groups in each bin is shown as filled violet circles of the size,
which was proportional to the relative number.
These data are overplotted over the general diagram for all 14838 ARs as it is shown in Fig.\,\ref{fig:butt}.

As the cycle proceeds, large violet circles in Fig.\,\ref{fig:relbutt} show a general tendency to appear more frequently
and closer to the equator.
Noticeable crowding on them is observed near the oncoming solar minimum.
A visible excess of anti-Hale ARs is also found at the edges of the butterfly diagram.

\section{Concluding remarks}

The catalog BARVHL  (1989--2020) and SMWMO data (1925--1958) on verified anti-Hale ARs allowed us to jointly explore the data
for five completed SCs (16--18, 23, 24) and thus to overcome the limitations associated with poor statistics
of anti-Hale ARs.
We developed a routine to compile a synthetic cycle that takes into account the time and
latitude positions of the ARs.
We thus obtained the synthetic cycle consisting of 14838 ARs, including 367 anti-Hale ARs.
As a result, we  would like to draw the following conclusions.

(i) The synthetic cycle has a typical shape of a SCs with two typical maxima.
The anti-Hale ARs demonstrate overall correspondence with the cycle that shows their intrinsic
connection with the global dynamo.
Apparently, such ARs have the same origin rooted in the coherent toroidal flux system in the convection zone
as all other ARs.
This is consistent with generally accepted theoretical and phenomenological concepts of the solar dynamo
(the mean-field dynamo theory, the Babcock-Lieghton mechnism) and the results of recent observations \citep{Munoz-Jaramillo21}.
The presence of the pronounced peak in the anti-Hale ARs number in the second maximum might mean
the involvement of other, besides the global dynamo mechanism of the magnetic field generation.
However, the convective zone (where the global dynamo operates) is a highly nonlinear dissipative medium
and the complexity of physical processes occurring there hinder further clarification about the probable mechanisms.
%The fluctuation dynamo seems to be plausible as such a mechanism \citep{Sokoloff15, Abramenko18, Zhukova20GA}.

(ii) The relative number of anti-Hale ARs in the synthetic cycle is mostly cycle-independent,
excluding the years of the solar minimum.
The peak in the anti-Hale ARs percentage in the minimum was also found for the coupling of two identical
synthetic cycles (we performed such a coupling with overlapping interval of two years to simulate an adjunction of cycles).
This result repeats the pattern reported for sequences of SCs 15-17, 20-24 by  \citet{McClintock14, Sokoloff15}.
\citet{McClintock14} associated this phenomenon with a possible misalignment of the magnetic and
heliographic equators, and this will be discussed below in more detail.
\citet{Sokoloff15} suggested the increasing influence of small-scale processes, which become noticeable
when the effect of the global dynamo is weakening.
Our results do not contradict their hypothesis.

(iii) The time-latitude diagram for the synthetic cycle shows that the anti-Hale ARs are spread nearly uniformly
over the diagram.
%If we presume that the appearance of such ARs is associated with \textbf{the fluctuations of the magnetic field
%(as it was assumed, for instance, in the probabilistic model by \citet{Sokoloff15})
%then such a distribution might indicate their ubiquity.}
%Although this fact might seem obvious, we emphasize it here because it is underlying
%for the mean-field dynamo theory \citep{Moffatt78, Krause80}.
%(iv)
The time-latitude distribution of the relative number of anti-Hale ARs shows
a number of features.
A pronounced excess of reverse-polarity groups is found near the equator.
At the same time, near the oncoming solar minimum, noticeable growth of the percentage is observed on various latitudes.
An excess in the fraction of anti-Hale ARs is also found along the edges of the butterfly diagram.
In addition, one can notice a week tendency for large circles in Fig.\,\ref{fig:relbutt}
to appear more often and closer to the equator as the cycle proceeds.

Most of the observed features occur at the maximum and during the declining phase of the cycle.
This period is known as the time of the polar field reversal and building the polar field that seeds the new cycle.
The increased number and percentage of anti-Hale ARs at that time show that the role of such groups
in the recovering of the global poloidal field may indeed be significant  \citep{Hazra17, Karak18, Mordvinov19, Mordvinov22}.
The enchanced percentage of the anti-Hale ARs on the diagram is observed when the toroidal magnetic field is weakend (the edges of the time-latitude diagram and area near oncoming solar minimum).
This growth might be related  to the presumably increased influence
of the magnetic field fluctuations, high nonlinearity and turbulence at magnetic flux tubes of ARs that emerges
through the convective zone \citep{LopezFuentes03, Knizhnik18, Abramenko21}.
Since a butterfly diagram is believed to be related with the the ``dynamo wave''
\citep{Parker55, Moffatt78, Krause80}, we could also associate an increase in the share of anti-Hale ARs
with some obstacles in a wave propagation.

It is also worth mentioning that the similar trends, as we found for the relative number
of anti-Hale ARs, were reported for helical characteristics of the magnetic field.
\citet{Zhang10, Xu15} found an increase of the average current helicity and twist at the edges
of the butterfly diagram and near the equator in SCs 23 and 24.
A possibility of the hemispheric rule reverse during the declining phase of each cycle was shown by
\citet{Miesch16} in their 3D MHD simulations of the convective dynamo.
Although a strong twisting of the magnetic field does not mean the formation of an anti-Hale AR in itself,
it generates the kink instability, which contributes to the formation of such ARs \citep{Knizhnik18}.

And finally, some remarks about the mutual location of the magnetic and heliopraghic equators should be made.
The importance of the anti-Hale ARs location and orientation relative to the equator was first
emphasized by \citet{Richardson48}.
\citet{Zolotova09}. Basing on a forty-years-observation data set, it was  estimated that the shift between the equators is around 4\textdegree.
Besides,  according to \citet{Obridko11}, the magnetic equator may have a complex shape.
The authors took into acount the mutual position of axial and magnetic dipoles.
\citet{McClintock14} suggested to explain the increase in the anti-Hale groups percentage
during a solar minimum as a result of a shift between the equators.
Indeed, an excess in the fraction of anti-Hale ARs in the equatorial region that was found
in this study supports their hypothesis.
However, as it was shown above, the increase in the anti-Hale ARs relative number
near the solar munimum occurs at various latitudes.
In this regard, we suppose that this increase may be due to more than a single reason apart from the equators' misalignment.

We would like to note that in the present work our primary focus was on the behavior of time-latitude distribution of the anti-Hale ARs
and our approach can be further used for the investigation of the physical properties of anti-Hale ARs themselves
e.g. \cite{Munoz-Jaramillo21}.

\section*{Acknowledgements}

We are grateful to Dr. Naga Varun for improving English.
This study was completed with the support of the Russian Science Foundation (project 18-12-00131).

%%%%%%%%%%%%%%%%%%%%%%%%%%%%%%%%%%%%%%%%%%%%%%%%%%
\section*{Data Availability}

%Data Availability Statements provide a standardised format for readers to understand the availability of data underlying the research results described in the article. The statement may refer to original data generated in the course of the study or to third-party data analysed in the article. The statement should describe and provide means of access, where possible, by linking to the data or providing the required accession numbers for the relevant databases or DOIs.
SOHO is a project of international cooperation between ESA
and NASA. The SDO/HMI data are available by courtesy of
NASA/SDO and the AIA, and HMI science teams.
The other data underlying this article are available at \citet{Crao19892018, DPD, SMWMO, USAF}.



\bibliographystyle{mnras}
\bibliography{Zhukova_bibl} % if your bibtex file is called example.bib

%%%%%%%%%%%%%%%%%%%%%%%%%%%%%%%%%%%%%%%%%%%%%%%%%%

% Don't change these lines
\bsp	% typesetting comment
\label{lastpage}
\end{document}